\documentclass[%
 reprint,
 amsmath,amssymb,
 aps,
]{revtex4-2}
\usepackage{subfigure}
\usepackage{graphicx}
\usepackage{dcolumn}
\usepackage{bm}
\usepackage[normalem]{ulem}
\usepackage{hyperref}
\usepackage{xcolor}

\begin{document}

\title{Defect liquids in a weakly imbalanced bilayer Wigner Crystal}
\author{Zekun Zhuang}
\author{Ilya Esterlis}
\affiliation{Department of Physics, University of Wisconsin-Madison, Madison, Wisconsin 53706, USA}

\date{\today}

\begin{abstract}
In a density-imbalanced bilayer Wigner crystal, where the ratio of electron densities in separate layers deviates slightly from unity, defects spontaneously form in one or both layers in the ground state of the system. Due to quantum tunneling, these defects become mobile and the system becomes a defect liquid. Motivated by this idea, we numerically study the semiclassical energetics of individual and paired point defects in the bilayer Wigner crystal system. We use these results in combination with a simple defect model to map the phase diagram of the defect liquid as a function of electron density and interlayer distance. Our results should be relevant for present experimental bilayer Wigner crystal systems. 
\end{abstract}

\maketitle

\section{Introduction}

Recent experiments across a variety of two-dimensional electron gas (2DEG) systems have provided compelling evidence for the existence of Wigner crystal (WC) phases at low electron densities \cite{Smolenski2021,Zhou2021,Falson2022,Hossain2020,Sung2023,Xiang2024,Tsui2024,Regan2020,Li2021,Regan2022}. Among these the bilayer WC \cite{Zhou2021,Manoharan1996,wang2007,wang2012,DOVESTON2002} -- formed in two Coulomb-coupled equal density electronic layers separated by a distance $d$ -- is an especially promising platform for investigating properties of the electron solid due to the possibility of realizing different WC lattice geometries with varying interlayer coupling strength \cite{Vilk1984,*Vilk1985,Falko1994,Narsimhan1995,Esfarjani1995,Goldoni1996,Goldoni1997,Samaj2012,rapisarda1996,*rapisarda1998} and the enhanced stability of the bilayer crystal to higher electron densities \cite{swierkowski1991,Goldoni1997,rapisarda1996,*rapisarda1998}. 

\begin{figure}
    \centering
    \includegraphics[width=0.5\textwidth]{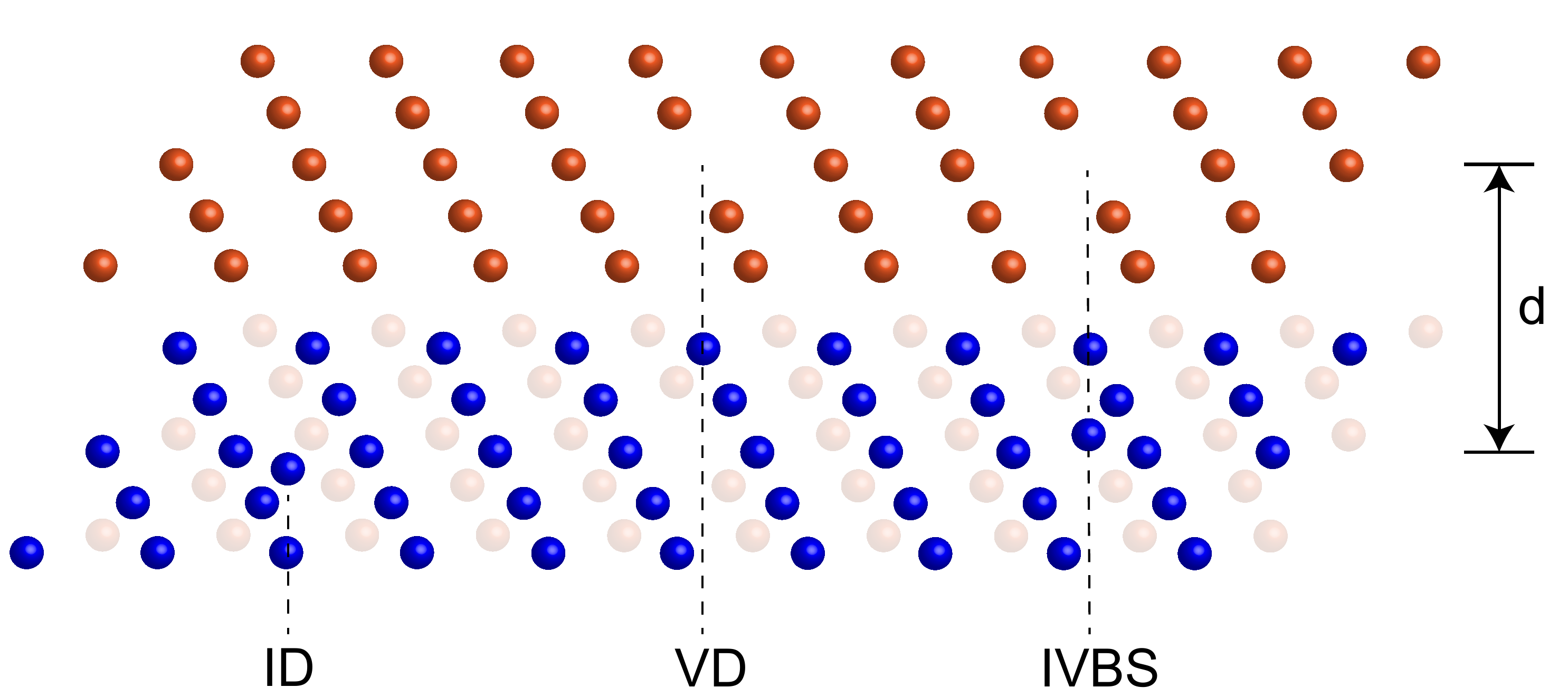}
    \caption{Three types of defects -- interstitial defect (ID), vacancy defect (VD), and interstitial-vacancy bound state (IVBS) -- in a imbalanced bilayer Wigner crystal. Lighter points indicate projection of electron positions from the top layer onto the bottom layer.}
    \label{fig:DefSchem}
\end{figure}

In any realistic bilayer WC system there will be some, possibly small, density mismatch $\delta n$ between electron densities in the two layers. Denoting the layer densities $n_1 = n+ \delta n$ and $n_2 = n$, a slight imbalance $|\delta n| \ll  n$ will lead to the introduction of a small concentration of defects in the ground state of the bilayer WC (if the interlayer coupling is not too weak), as shown in Fig. \ref{fig:DefSchem}. Even a dilute concentration of such defects may have outsized effects on the properties of the system. For example, in the monolayer WC it was recently found that tunneling barriers for interstitials and vacancies are small, being significantly reduced compared to tunneling barriers associated with ring-exchange processes \cite{Kim:2022,Kim2024}. The dynamics of these defects leads to kinetic magnetism with energy scales much larger than those associated with ring-exchange \cite{Kim:2022}, as well as the possiblity of a self-doping instability of the monolayer WC to a metallic crystal state \cite{Kim2024}. If the tunneling barriers are similarly small in the bilayer WC, then the dynamics of a dilute concentration of defects in a weakly imbalanced bilayer may have important implications for the magnetism and transport properties of the system. 

In the present paper we numerically investigate the energetics of interstitials and vacancies in the bilayer WC, as well as interstitial-vacancy bound states between layers. We determine the lowest energy defects as a function of layer density $n$ and interlayer separation $d$ in the semiclassical approximation, which includes the classical electrostatic and zero-point vibrational energies of the defects. Combining these results with a simple defect model we determine the ``ground state defect" phase diagram of the system. Even at zero temperature (and in a clean system), such defects will in principle delocalize via quantum tunneling, leading to the existence of different metallic ``defect liquid" states \cite{Andreev1969,Falakshahi2005,Kim2024}. 

The paper is organized as follows: In Sec. \ref{Sec:Formulation}, we briefly introduce the basics of the (balanced) bilayer WC and then describe the defect model of the imbalanced bilayer WC. We also describe our numerical methods. We present our results in Sec. \ref{Sec:results} and conclude in Sec. \ref{Sec:Conclusions}.
\section{Formulation}\label{Sec:Formulation}
\subsection{Balanced bilayer Wigner Crystal}
\begin{figure}
    \centering
    \includegraphics[width=0.5\textwidth]{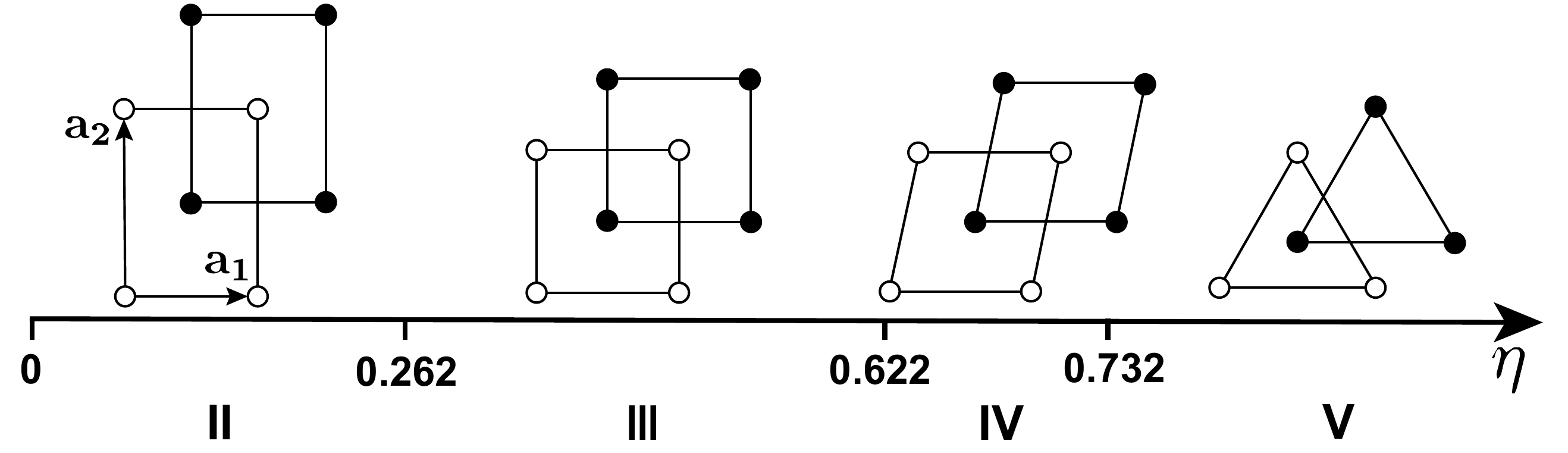}
    \caption{The phase diagram of the bilayer Wigner crystal in the classical limit $r_s\rightarrow \infty$. Circles with different colors denote different layers. Note that phase I is not shown as it only exists at $\eta=0$ and can be regarded as a special case of phase II with $|\mathbf{a_2}|=\sqrt{3}|\mathbf{a_1}|$.}
    \label{fig:BLWCevolution}
\end{figure}
The Hamiltonian of the bilayer 2DEG is given by
    \begin{align}\label{Hamiltonian}
    H = \sum_{i}\frac{\mathbf p_{i}^2}{2m} &+\frac{1}{2} \sum_{i\neq j} \frac{e^2}{ \sqrt{|\mathbf r_{i} - \mathbf r_{j}|^2 + d_{ij}^2}}. 
    \end{align}
Here $m$ is the effective electron mass, $\mathbf p_{i}$ and $\mathbf r_{i}$ refer to the in-plane momentum and lateral coordinate of the $i$th electron, and $d_{ij}$ is the distance between electrons $i$ and $j$ along the vertical direction, which equals $d$ for electrons in different layers and zero if they are in the same layer. A charge-neutralizing background is implicitly assumed to make the energy convergent. When the electron densities in the two layers are equal,  $n_1=n_2 = n$, the system is characterized by two dimensionless parameters: $r_s=a/a_B$ and $\eta=d\sqrt{n}$, where $a=1/\sqrt{n \pi}$ denotes the average distance between electrons in each layer and $a_B=\hbar^2 /me^2$ is the Bohr radius. 

In the extreme dilute limit $r_s\rightarrow \infty$, the potential term dominates over the kinetic term in Eq.~\eqref{Hamiltonian} and the phase diagram as a function of $\eta$ is completely determined by minimizing the classical electrostatic energy. It is known \cite{Vilk1984,*Vilk1985,Falko1994,Narsimhan1995,Esfarjani1995,Goldoni1996,Goldoni1997,Samaj2012} that there are five different phases in this limit, as depicted in Fig. \ref{fig:BLWCevolution}. In order of increasing $\eta$, the phases are: one-component hexagonal (I), staggered rectangular (II), staggered square lattice (III), staggered rhombic (IV), and staggered hexagonal (V) state. Phases I-IV are connected via sequential second-order phase transitions, while the transition from IV-V transition is first-order.

\begin{figure}
    \subfigure[]{\label{fig:A1}
    \includegraphics[width=0.48 \textwidth]{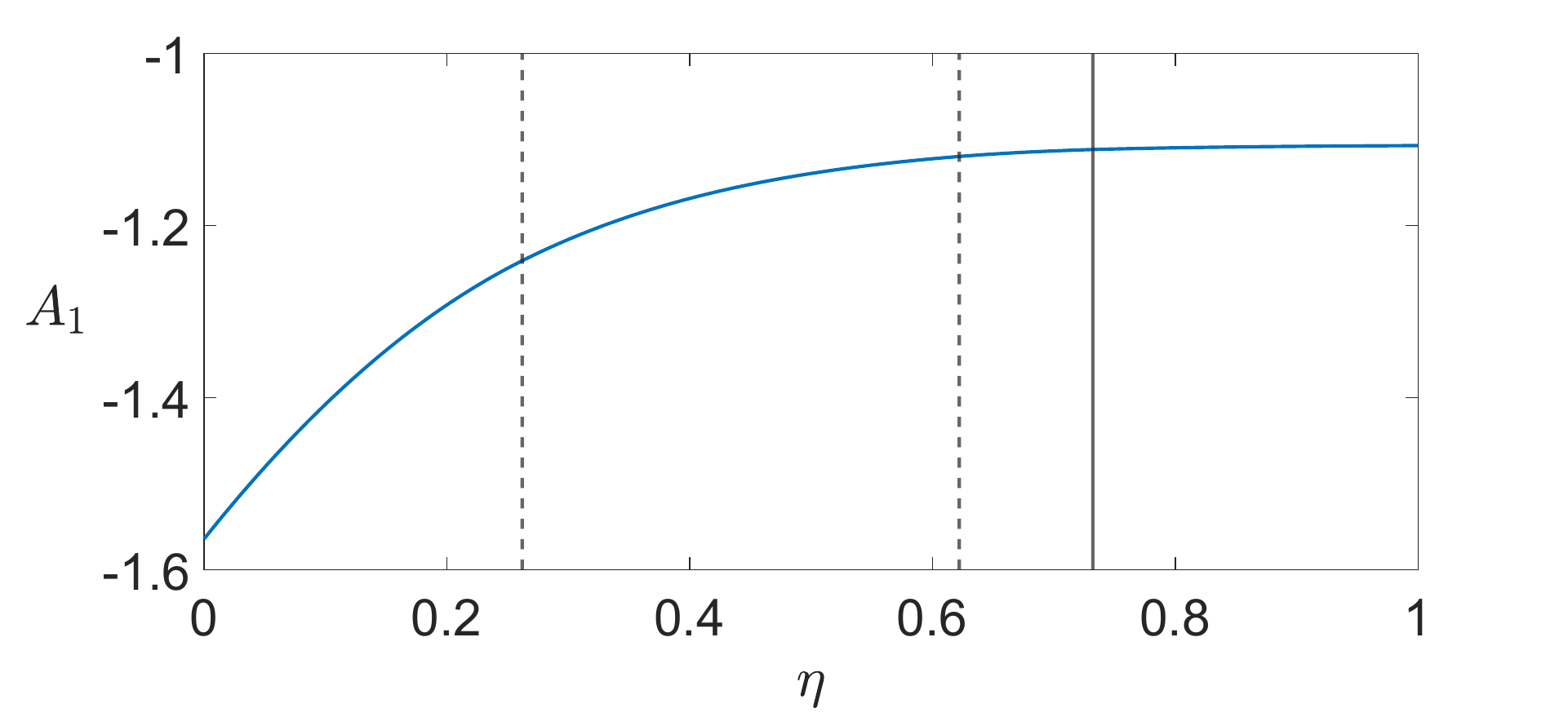}}
    \subfigure[]{\label{fig:A15}
    \includegraphics[width=0.48 \textwidth]{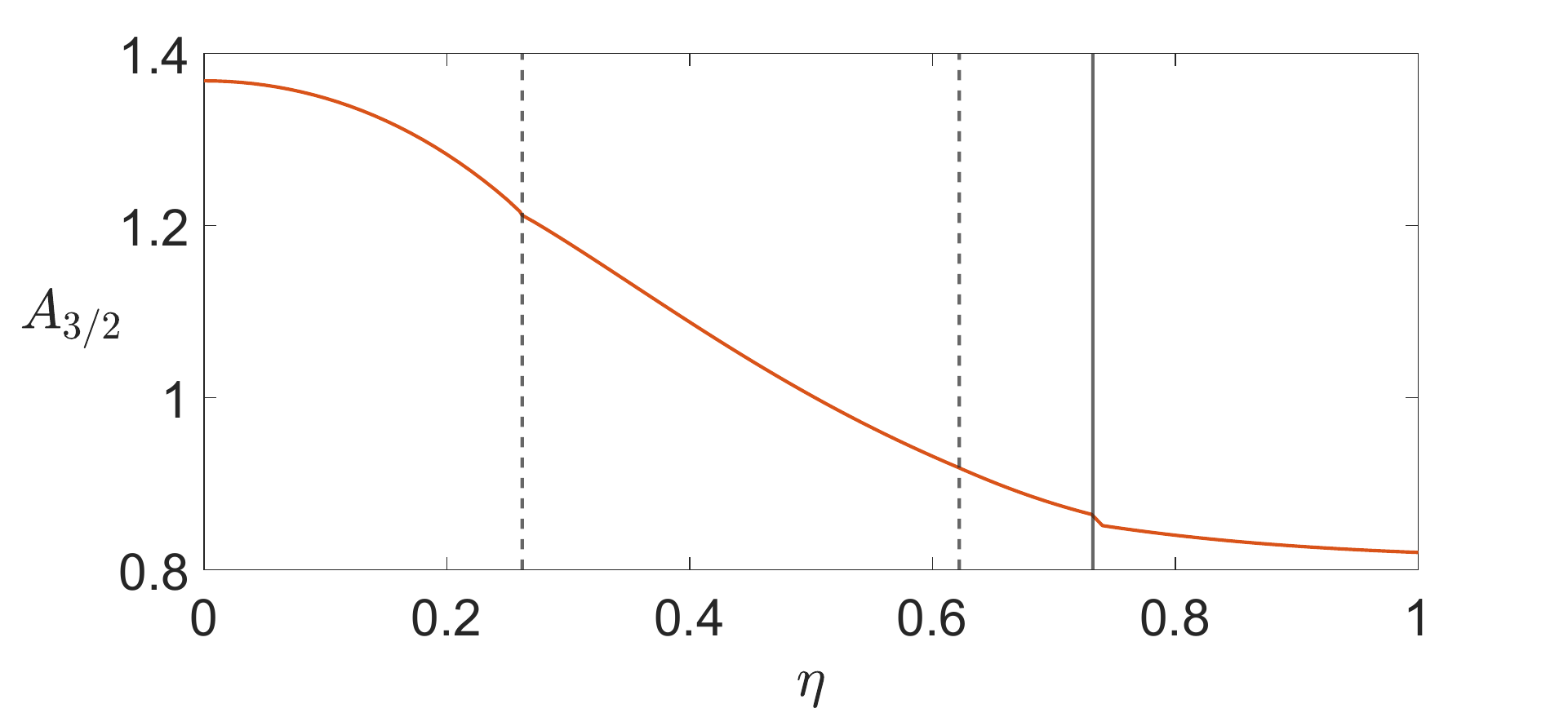}}    
    \caption{Coefficients $A_1$ and $A_{3/2}$ as a function of $\eta$. The vertical solid (dashed) line denotes the first(second)-order phase transition (same for Figs. \ref{fig:B}, \ref{fig:C1D1} and \ref{fig:phasediagram}).}
   \label{fig:A} 
\end{figure}

The leading quantum correction to the ground state energy comes from zero-point motion of the WC phonons. Expressed in the Hartree energy unit $\hbar^2/ma_B^2$, the semiclassical expansion for the ground state energy per particle takes the form \cite{Bonsall1977,Tanatar1989}:
\begin{equation}
    \epsilon (r_s)=\frac{A_1}{r_s}+\frac{A_{3/2}}{r_s^{3/2}},\label{def:A}
\end{equation}
where the first and second terms correspond respectively to the electrostatic and vibrational energies. The dimensionless coefficients $A_1$ and $A_{3/2}$ can be computed numerically using the Ewald technique discussed in Sec. \ref{sec:numerical} and Appendix \ref{app:Ewald} (see also Ref. \cite{Goldoni1996}). Their dependencies on $\eta$ across the various bilayer WC geometries are shown in Fig. \ref{fig:A}.

\subsection{Defect model of the imbalanced bilayer Wigner crystal }\label{Sec:DefModel}
When the bilayer crystal is weakly imbalanced $ |\delta n|\ll n$, defects spontaneously form in one or both layers due to the mismatching lattice constants. We will only consider the formation of point defects, away from which the lattices in both layers are only slightly distorted. We do not consider the situation in which the two layers are ``depinned" -- i.e., that they may have different lattice constants and form moir\'e-like patterns -- which may occur at large enough $\eta$ via an Aubry-type transition \cite{AUBRY1983,Mandelli2017,Huang2022}. The elementary point defects we consider are the vacancy defect (VD) and interstitial defect (ID), corresponding respectively to removing or adding one electron from or to a layer. The density of defects in each layer satisfies the constraint $\nu_{1,i}-\nu_{1,v}-\nu_{2,i}+\nu_{2,v}=n_1-n_2 = \delta n$, where $\nu_{\ell,i(v)}$ denotes the density of IDs (VDs) in layer $\ell$.

We adopt a simplified model of the imbalanced bilayer based on the following assumptions: (i) The VD and ID in the same layer attract each other strongly, so that an intralayer ID-VD pair always annihilates. (ii) If the VD and ID in different layers attract, they form an interstitial-vacancy bound state (IVBS). If they repel, their repulsive interaction can be neglected in the limit of dilute defects. (iii) There is no other interaction among defects, including the IVBS. (iv)  The system can lower its energy by annihilating interlayer pairs of like defects while simultaneously adjusting the lattice constant. Assumptions (i) and (ii) are based on the observation that a perfect lattice has the lowest energy and that an interstitial electron tends to be locked to the vacant site in the opposite layer to minimize the electrostatic energy. Assumption (iv) is valid at large $r_s$, where creating defects is always energetically unfavorable. However, at small $r_s$ this assumption may break down as the vibrational energy dominates over the electrostatic energy, potentially making the proliferation of defects favorable \cite{Cockayne1991}. We do not consider this possibility. Assumption (iii) is less justified, as it has been shown that defects in a monolayer WC may attract each other via a Wigner-crystal phonon-induced electron attraction \cite{Candido2001}. We assume that over most of the parameter space, the energy scale of this interaction is small compared to the other energy scales and may be neglected. 

\begin{figure}
    \centering
    \includegraphics[width=0.5\textwidth]{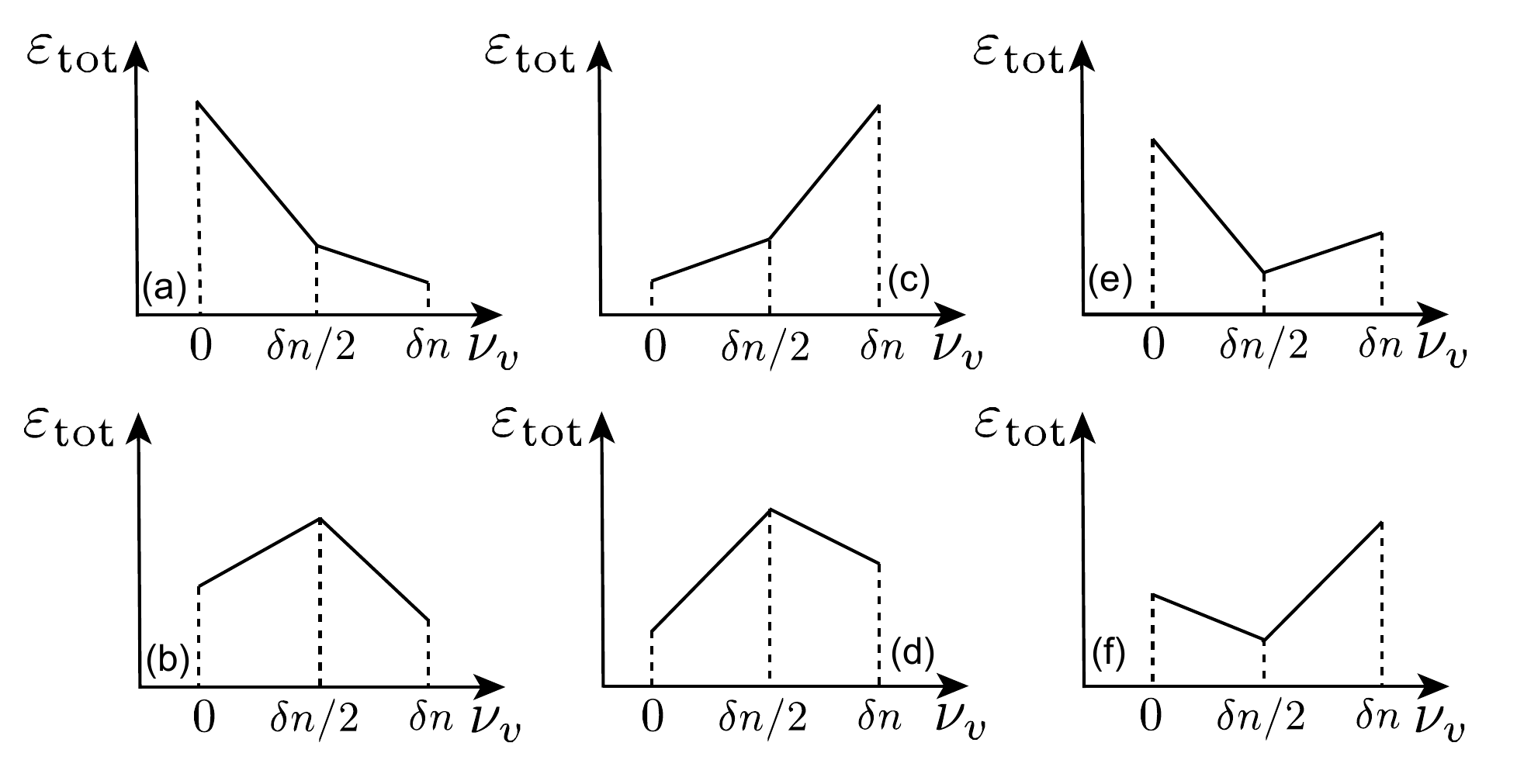}
    \caption{Schematic plot of the possible dependencies of the energy $\varepsilon_\text{tot}$ on the VD density $\nu_{v}$ for (a, b) VD, (c, d) ID, and (e, f) IVBS state. }
    \label{fig:Evsnu}
\end{figure}
With the above assumptions in mind and assuming $\delta n=n_1-n_2>0$, the ground state of an imbalanced bilayer WC is described by a state having IDs (VDs) in the first (second) layer with density $\nu_{i}$ $(\nu_{v})$, subject to the constraint $\delta n = \nu_i + \nu_v$. Due to the attraction, equal amounts of IDs and VDs form IVBSs of density $\nu_{iv}=\text{min}\{\nu_{i},\nu_{v}\}$, leaving the rest of the IDs or VDs unpaired. As demonstrated below, there are three possible different defect liquid ground states: $\nu_i = \delta n,~\nu_v = 0$ (ID liquid), $\nu_i = 0,~\nu_v = \delta n$ (VD liquid), or $\nu_i = \nu_v = \delta n /2$ (IVBS liquid). 
\begin{figure*}[!ht]
    \subfigure[]{\label{fig:Int}
    \includegraphics[width=1 \textwidth]{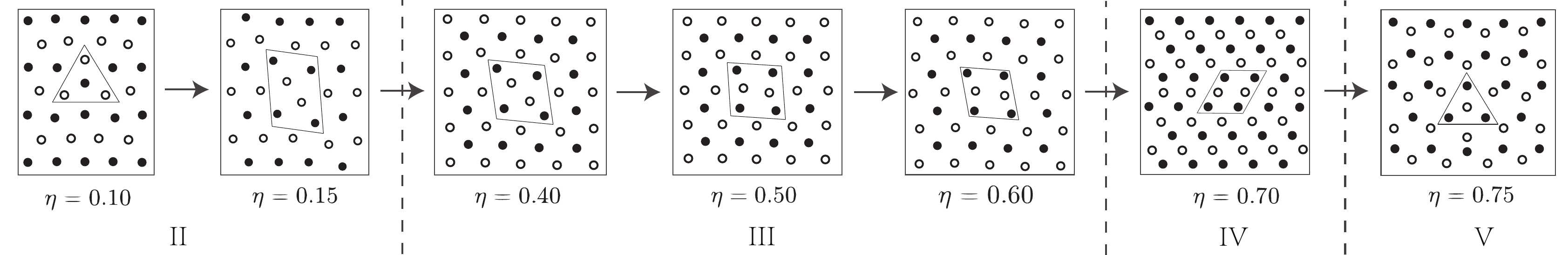}}
    \subfigure[]{\label{fig:Vac}
    \includegraphics[width=1 \textwidth]{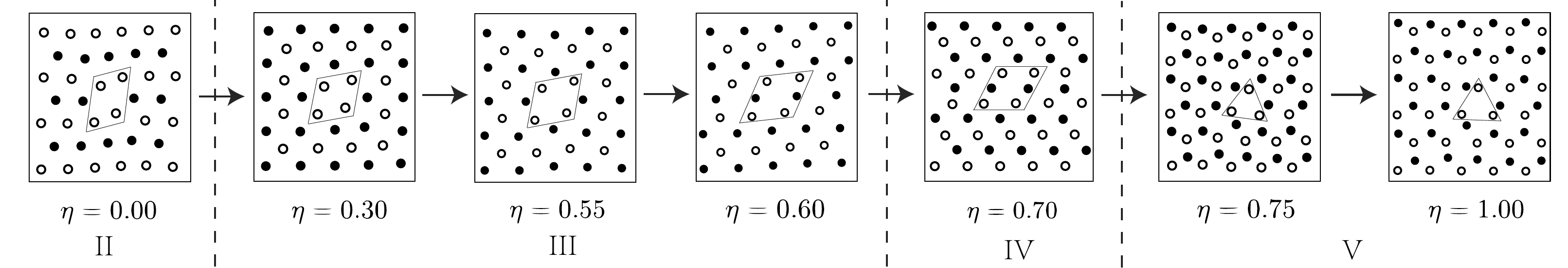}}    
    \subfigure[]{\label{fig:IV}
    \includegraphics[width=1 \textwidth]{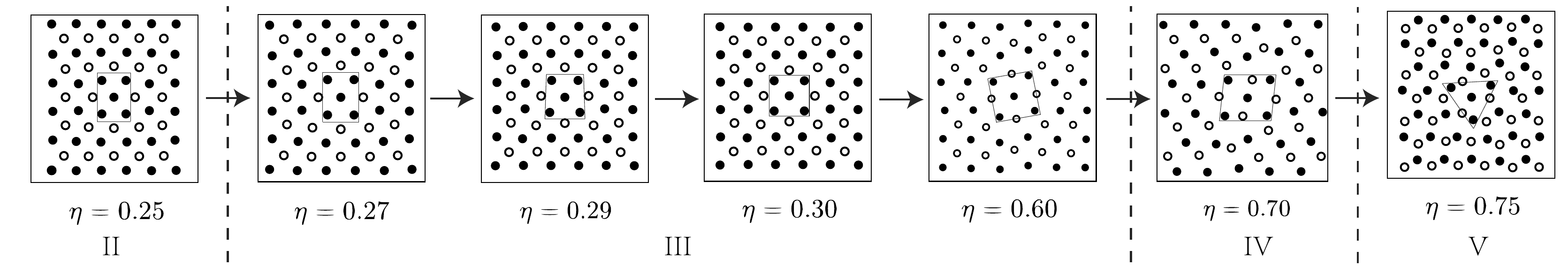}}
    \caption{Representative configurations for (a) interlayer defect (ID) (b) vacancy defect (VD) (c) interstitial-vacancy bound state (IVBS) defect for different values of the dimensionless interlayer spacing $\eta=d\sqrt n$. Black and white markers correspond to electrons in different layers. In (a), the interstitial electron is placed in the white layer. In (b), the vacancy is in the black layer.}
   \label{fig:Config} 
\end{figure*}

To calculate the energy of each defect liquid state, we first define the chemical potential of a single defect:  
\begin{equation}
     \mu_D=E_D-N \epsilon(r_s) \equiv \frac{B_1^{D}}{r_s}+\frac{B_{3/2}^{D}}{r_s^{3/2}}. \label{def:B}
\end{equation}
Here $E_D$ is the energy of the system with a single defect of type $D$ ($D=v, i, iv$ correspond to \text{VD}, \text{ID}, \text{IVBS} respectively), $\epsilon(r_s)$ is the ground state energy per particle of the perfect crystal defined in Eq.~\eqref{def:A}, and $N$ is the total number of electrons in the absence of the defect. Similar to Eq. \eqref{def:A}, we have defined coefficients $B^D_1$ and $B^D_{3/2}$ associated to the leading terms in the semiclassical expansion of the defect chemical potential (electrostatic and zero-point vibrational energies, respectively). 

We now consider the energy density $\varepsilon_D$ -- as opposed to energy per particle, as in Eq. \eqref{def:A} -- of the various defect ground states to first order in $\delta n$. The energy density of the balanced bilayer per layer is denoted $\varepsilon_0(n)=\epsilon(r_s)/\pi r_s^2$ where $r_s=1/\sqrt{\pi n}$, and the associated chemical potential $\mu_0 = \partial \varepsilon_0 / \partial n$. For an ID, it is straightforward to show\footnote{To avoid confusion: in Eqs.~\eqref{eq:id}-\eqref{eq:ivbs}, $n$ and $\delta n$ inside all the paratheses should be understood as the arguments of the corresponding quantities.}
    \begin{equation}
            \varepsilon_i(n,\delta n) = 2\varepsilon_0(n) + \mu_i(n) \delta n \quad \text{(ID)}.
    \label{eq:id}
    \end{equation}
For a VD, the system can be regarded as having vacancies with concentration $\delta n$, on top of a balanced bilayer with electrons of density $n+\delta n$  per layer. Therefore, the energy density for VD is given by
    \begin{align}
          \varepsilon_v(n,\delta n ) &= 2\varepsilon_0(n+\delta n) + \mu_v(n+\delta n) \delta n \nonumber \\ 
          &\approx 2 \varepsilon_0(n) + [2\mu_0(n) + \mu_v(n)]\delta n \quad \text{(VD)}.
          \label{eq:vd}
\end{align}
Similarly, for the IVBS state, one obtains 
\begin{align}
     \varepsilon_{iv}(n,\delta n) &= 2\varepsilon_0(n+\delta  n /2) + \mu_{iv}(n+\delta n/2)\delta n / 2 \nonumber \\
     &\approx  2 \varepsilon_0(n) + [2\mu_0(n)+ \mu_{iv}(n)]\delta n / 2 \quad \text{(IVBS)}. 
    \label{eq:ivbs}
\end{align}

We define the single defect energy difference as
    \begin{align}\nonumber
        \Delta \varepsilon_s(n) &= \frac{\varepsilon_v(n, \delta n) - \varepsilon_i(n, \delta n)}{\delta n} \\\nonumber
        &= 2\mu_0(n) + \mu_v(n) - \mu_i(n)\\
        &\equiv \left(\frac{D_1}{r_s}+\frac{D_{3/2}}{r_s^{3/2}}\right) \label{def:D},
    \end{align}
where $D_1=3A_1+B_1^v-B_1^i$ and $D_{3/2}=7/2A_{3/2}+B_{3/2}^v-B_{3/2}^i$. 
The energy differences between the IVBS liquid state and the single defect states are
    \begin{align}
        \frac{\varepsilon_{iv}(n, \delta n) - \varepsilon_v(n, \delta n)}{\delta n} &= \frac 12 [ E^b_{iv}(n)- \Delta \varepsilon_s(n)], \label{Ediffiv1}\\
        \frac{\varepsilon_{iv}(n, \delta n) - \varepsilon_i(n, \delta n)}{\delta n} &= \frac 12 [E^b_{iv}(n) + \Delta \varepsilon_s(n)],\label{Ediffiv2}
    \end{align}
where we have defined the binding energy of the IVBS
\begin{equation}
    E^b_{iv}=\mu_{iv}-\mu_i-\mu_v \equiv \frac{C_1}{r_s}+\frac{C_{3/2}}{r_s^{3/2}}, \label{def:C}
\end{equation}
which is negative (positive) if the ID and VD in different layers attract (repel). 

Depending on the signs of Eqs. (\ref{def:D})(\ref{Ediffiv1}) and (\ref{Ediffiv2}), the total energy density as a function of $\nu_v$ has different possible dependences, which are shown schematically in Fig. \ref{fig:Evsnu}. Although we have only considered three specific cases with $\nu_v=0, \delta n/2, \delta n$, one can show that $\epsilon_\text{tot}$ changes linearly with $\nu_v$ when $0<\nu_v<\delta n/2$ or $\delta n/2 <\nu_v <\delta n$ in the dilute defect limit, so these three cases are indeed the only important ones. For the IVBS liquid to be the lowest-energy state, Eqs. (\ref{Ediffiv1}) and (\ref{Ediffiv2})  may be combined into the single condition $E^b_{iv} + |\Delta \varepsilon_s| < 0$. If $E^b_{iv} + |\Delta \varepsilon_s| > 0$, then  the ground state is an ID liquid for $\Delta \varepsilon_s > 0$, while for $\Delta \varepsilon_s <0$ the groud state is a VD liquid.

In the next section we outline the numerical procedure used to determine coefficients $B_1^D$ and $B_{3/2}^D$  in Eq. \eqref{def:B} for the various defects, from which coefficients $D_1$ and $D_{3/2}$ \eqref{def:D} and $C_1$ and $C_{3/2}$ \eqref{def:C} are obtained.

\subsection{Numerical procedure}\label{sec:numerical}
Our numerical calculations of the defect energies employ a supercell commensurate with the defect-free bilayer WC lattice for a given $\eta$. That is, if the lattice primitive vectors are $\mathbf a_1$ and $\mathbf a_2$, then we choose a supercell with primitive vectors $\mathbf c_1 = L \mathbf a_1$ and $\mathbf c_2 = L \mathbf a_2$ with $L$ an integer. A defect-free bilayer WC cell thus contains $N = 2L^2$ electrons.

 To obtain a VD configuration, we remove an electron from one layer and minimize the electrostatic energy. For an ID configuration, we add an electron to the center-interstitial position of the layer and then perform the minimization. The initial condition for the IVBS is set by moving an electron from one layer to the other at the same position. In doing this we implicitly assume that the basis vectors $\mathbf a_1$ and $\mathbf a_2$ as well as the structural phase boundaries are unaltered by the presence of defects. 

The electrostatic and vibrational energy of each configuration are calculated using the Ewald summation method (see the Appendix and Refs. \cite{Bonsall1977,Goldoni1996} for more details). We have considered supercell sizes up to $L=27$, and have verified that the defect configurations reported are dynamically stable \footnote{For the ID in part of phase IV we find there may be one unstable phonon mode and, when the relative distance between sites changes randomly by $0.1\%$, the corresponding phonon frequency fluctuates around zero while all other real phonon frequencies are almost unaffected. Since this is the only such configuration we can find, we suspect that the unstable phonon mode may be an artifact due to the numerical precision, and this defect is marginally stable.}. The value of $E_D$ in Eq. \eqref{def:B} is obtained by performing an extrapolation to the thermodynamic limit $L=\infty$. We find that the finite-size correction to the electrostatic part $B_1^D$ scales as $L^{-2}$ in all cases (see Appendix for an example), which is inconsistent with the early prediction by Fisher \textit{et al.} \cite{Fisher1979}, but agrees with the later observation by Cockayne and Elser \cite{Cockayne1991}. For the vibrational part $B_{3/2}^D$, we find that the form $L^{-3}$ suggested in Ref. \cite{Cockayne1991} fits well for most of the data and we assume this scaling relation in this work. 

\section{Results}\label{Sec:results}

\subsection{Defect configurations}

Figure \ref{fig:Int} shows representative ID configurations for increasing $\eta$. At small $\eta$, the ID configuration is smoothly connected to the single-layer centered-interstitial defect (SLCID) at $\eta=0$. For $\eta \gtrsim 0.11$, a new type of ID with two-fold rotational symmetry has lower energy. This defect may be regarded as a finite-$\eta$ extension of the single-layer edge-interstitial defect (SLEID), which is dynamically unstable at $\eta=0$ \cite{Fisher1979,Cockayne1991}. The shape of the ID continuously changes when $\eta$ further increases until a first-order transition to phase V occurs. In the limit $\eta\rightarrow \infty$, the ID configuration in phase V corresponds to a stack of triangular WCs with a SLCID.

In comparison, for a wide range of $\eta \geq 0$ the VD is continuously connected to the single-layer vacancy defect (SLVD), which has two-fold rotational symmetry \cite{Cockayne1991,Kim2024}; we  refer to this defect configuration as SLVD-2 (see Fig. \ref{fig:Vac}). In phase III a new VD structure emerges for $\eta \gtrsim 0.6$ due to the softening of a phonon mode and persists to phase IV. The structure of the VD then drastically changes across the discontinuous phase transition from phase IV to phase V.
Interestingly, the VD in phase V is not a stack of a perfect layer with an SLVD-2 as they have different rotational symmetries; instead, the structure may be regarded as being composed of a vacancy defect with three-fold symmetry. In fact, in the single-layer limit $\eta=0$, this $C_3$-symmetric vacancy defect, SLVD-3, is a dynamically stable configuration with energy slightly higher than that of SLVD-2, as noted in Ref. \cite{Price1991}.

The evolution of the IVBS is illustrated in Fig. \ref{fig:IV}. The configuration at $\eta=0$ is equivalent to a perfect single-layer WC, which then continuously deforms until phase V is reached. Interestingly, even though the base lattice has 4-fold rotational symmetry in phase III, in the narrow range $0.26 \lesssim \eta \lesssim 0.29$, the defect has lower symmetry. Around $\eta\approx 0.59$, a phonon mode is softened, leading to the rotation of the defect around its center. In phase V, the IVBS can be regarded as a direct stacking of SLCID and SLVD-3.
\begin{figure}[!t]
    \subfigure[]{\label{fig:B1}
    \includegraphics[width=0.45 \textwidth]{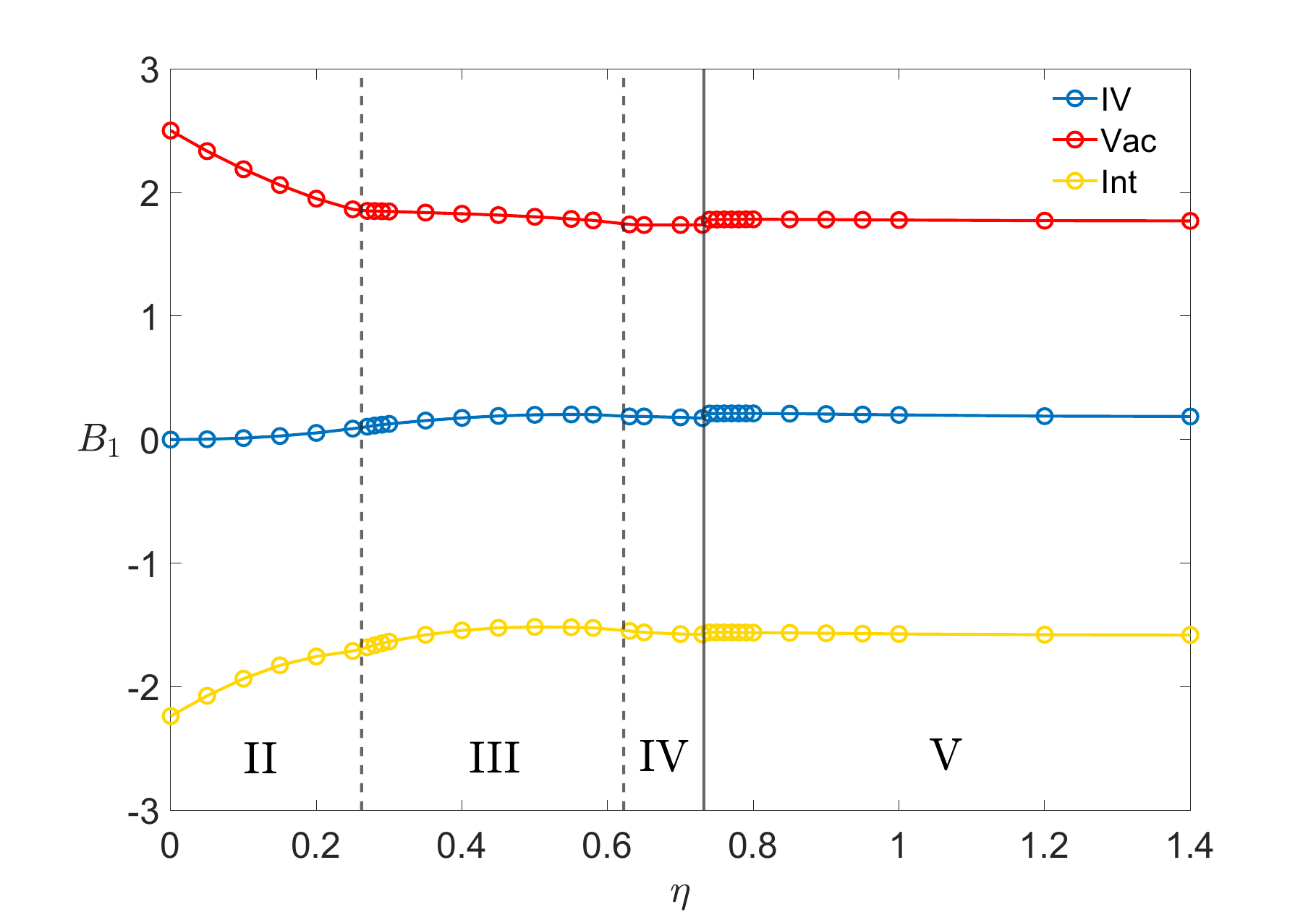}}
    \subfigure[]{\label{fig:B15}
    \includegraphics[width=0.45 \textwidth]{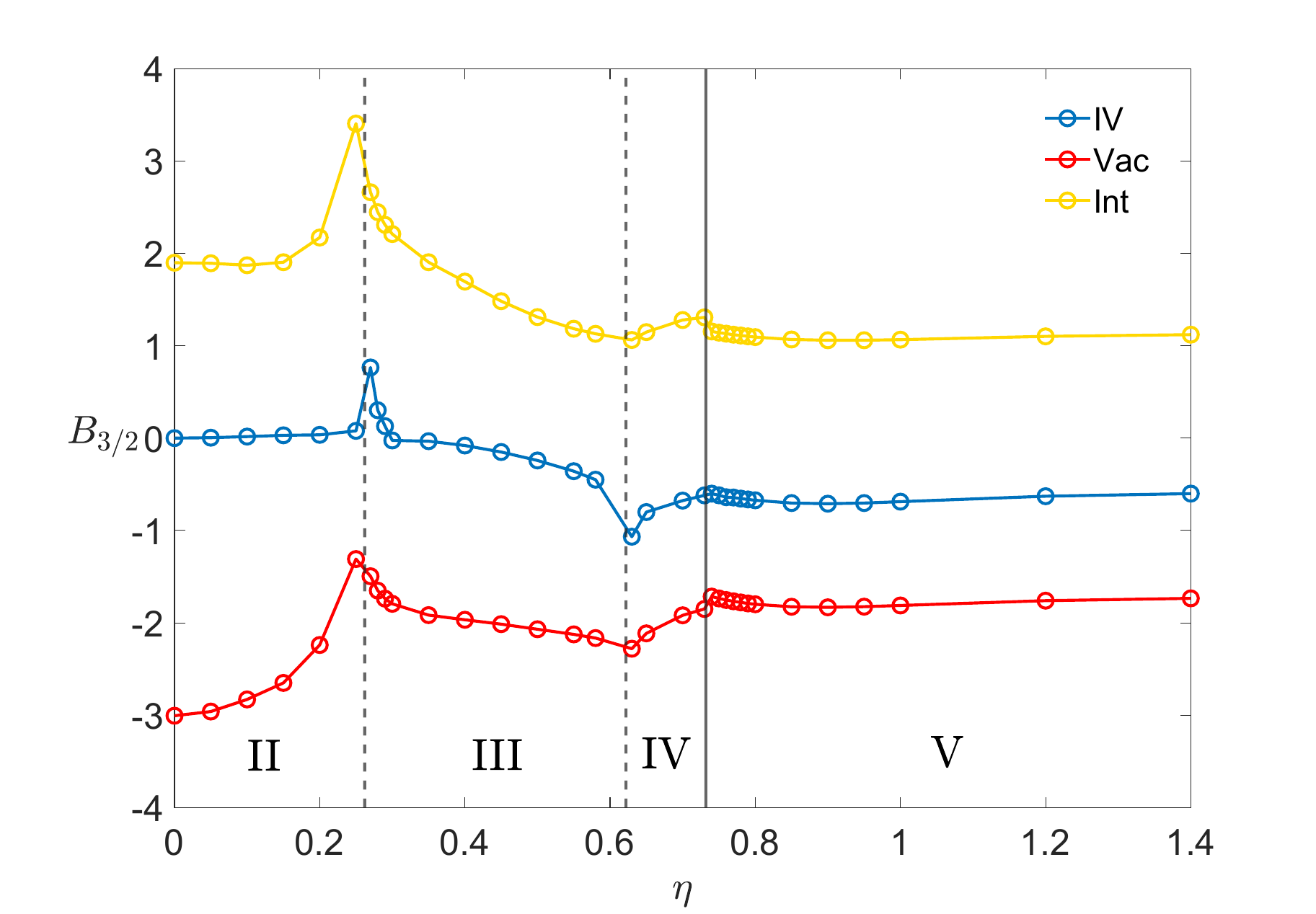}}    
    \caption{Chemical potential coefficients $B_1$ and $B_{3/2}$ versus $\eta$.}
   \label{fig:B} 
\end{figure}
\subsection{Phase diagram}

\begin{figure}[!ht]
    \centering
    \includegraphics[width=0.5\textwidth]{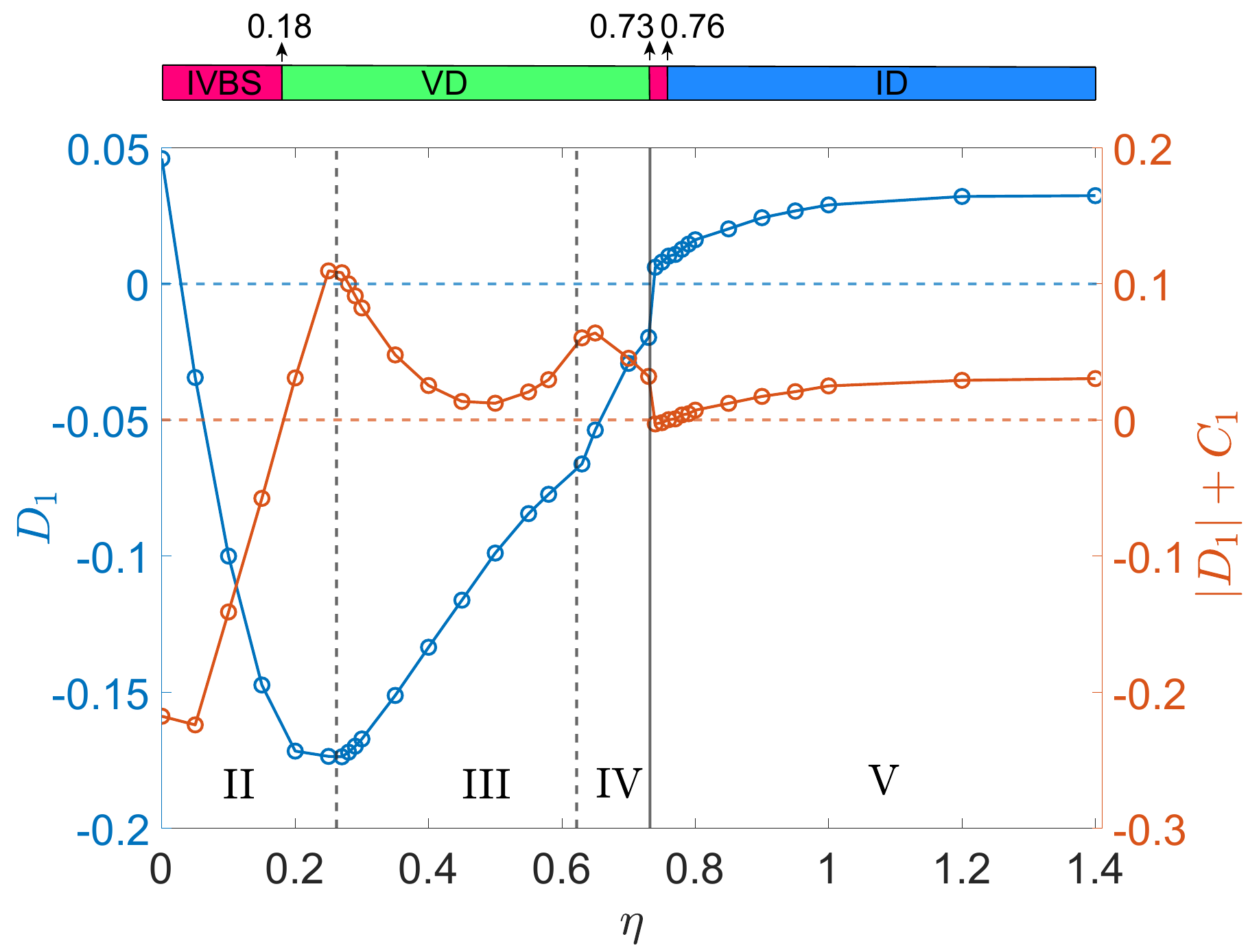}
    \caption{$D_1$ and $|D_1|+C_1$ versus $\eta$. The phase diagram in the classical $r_s \to \infty $ limit is indicated on top.}
    \label{fig:C1D1}
\end{figure}

In this subsection we present results for the ground state defect phase diagram as a function of $\eta$ and $r_s$. The results are based on our numerical calculations of the 
coefficients $B^D_1$ and $B^D_{3/2}$ determining the defect chemical potential Eq.~\eqref{def:B}. These coefficients are plotted in Fig.~\ref{fig:B}. We attribute the kinks in $B^D_{3/2}$ near the second-order phase transitions to strong effects of the defects on the soft phonon modes.

In the classical limit $r_s\rightarrow \infty$ where the zero-point contribution to the energy may be ignored, the defect ground state is determined just by the interlayer distance $\eta$. In this limit the criteria discussed in Sec.~\ref{Sec:DefModel} determining the defect configuration become the following: If $|\Delta \varepsilon_s| + E^b_{iv} = |D_1| + C_1 > 0$, then the the system is a VD liquid for $\Delta \varepsilon _s = D_1 <0$ and an ID liquid for $\Delta \varepsilon_s = D_1 >0$. If $|\Delta \varepsilon_s| = |D_1| + C_1 < 0$ then the system is an IVBS liquid.

In Fig. \ref{fig:C1D1}, we plot $D_1$ and $|D_1|+C_1$ as a function of $\eta$. The IVBS is favored at small $\eta$, as it corresponds to having no defects in the single-layer limit $\eta=0$. When $\eta\gtrsim 0.18$, the energy gain by converting an ID to a VD exceeds the binding energy of the IVBS, and the system becomes a VD liquid in the ground state, which persists until the first-order transition to phase V. In phase V, the system is an ID liquid except for a narrow range $0.73\lesssim \eta\lesssim 0.76$ where the IVBS is favored.

\begin{figure}[!b]
    \centering
    \includegraphics[width=0.45\textwidth]{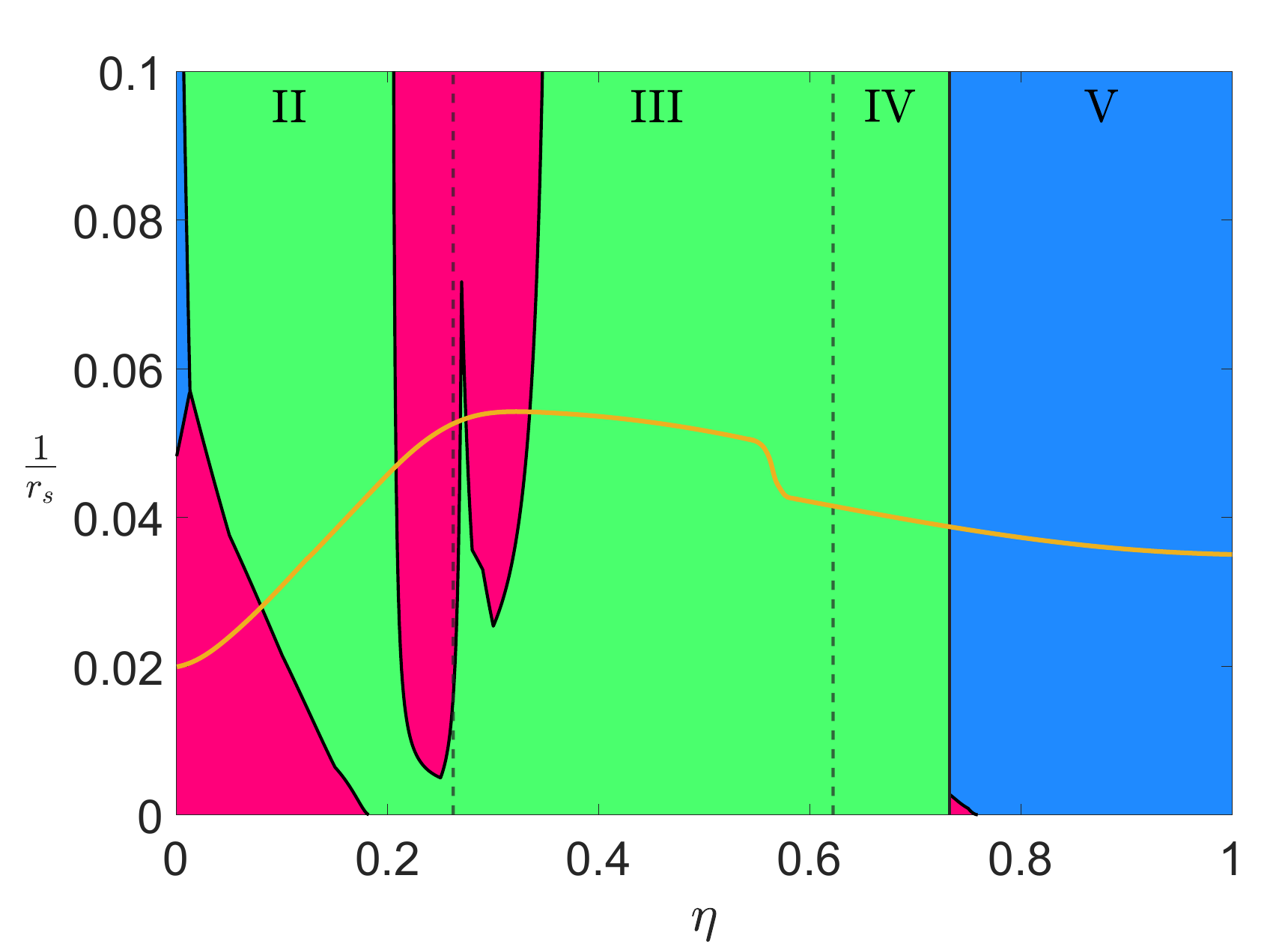}
    \caption{The phase diagram of the defect liquid, where red, green and blue regions denote the IVBS, VD and ID respectively. We interpolate the data in Fig. \ref{fig:B} to obtain a smooth phase boundary. The yellow line indicates the solid-homogeneous liquid transition extracted from the data in Ref. \cite{rapisarda1996} for a balanced bilayer.}
    \label{fig:phasediagram}
\end{figure}

The defect phase diagram at finite $r_s$ is obtained by taking into account both the classical electrostatic and zero-point vibrational energies. Our results are shown in Fig. \ref{fig:phasediagram}, in which the solid-homogeneous liquid transition (yellow line) is sketched based on the results from Ref. \cite{rapisarda1996,*rapisarda1998}\footnote{The transition line was determined in \cite{rapisarda1996,rapisarda1998} for a balanced bilayer; a non-zero (but small) defect density will likely lead to correspondingly small modifications in the phase boundary.}. We note that anharmonic effects beyond the harmonic approximation may become relevant at smaller $r_s$, which are out of the scope of this work. Quantum zero-point fluctuations either enhance or suppress the binding energy $E^b_{iv}$ and the chemical potentials $\mu_i$, $\mu_v$, thereby leading to diverse phases at finite $r_s$. We find the possiblity of a VD to IVBS transition with increasing density (decreasing $r_s$) near the II-III phase boundary, as well as an IVBS to VD transition with increasing density for $\eta \lesssim 0.2$. The kink in the phase boundary near the II to III structural transition can be traced back to the anomalous behavior of $B^D_{3/2}$ (see Fig.~\ref{fig:B15}). 

\section{Conclusion}\label{Sec:Conclusions}
In this paper we have numerically investigated the energetics of individual and paired point defects in a weakly imbalanced bilayer WC. Utilizing a simplified defect model, we have mapped out the ground state defect phase diagram as a function of interlayer distance and electron density in the semiclassical approximation. We have found three possible types of defect liquid states -- interstitial defect (ID) liquid, vacancy defect (VD) liquid, and interstitial-vacancy bound state (IVBS) liquid -- all of which exist in a wide range of parameters (see Fig. \ref{fig:phasediagram}).

Defect tunneling process should have a number of consequences for the transport and magnetic properties of the weakly imbalanced bilayer WC. While the perfect bilayer WC is an insulator, the charged ID and VD in the imbalanced system will conduct current when a voltage is applied to both layers (so long as the system is sufficiently clean that the defects are not localized). The IVBS is charge-neutral and would not respond to a uniform field, but could be detected by measuring drag resistivity \cite{Levchenko2016}. Noise spectroscopy \cite{Dolgirev2024} may also be a useful probe of the defect liquid phases we have described, as we expect the metallic character of the itinerant defects to lead to an enhanced current noise relative to the balanced bilayer.

Defect dynamics may also influence the magnetism of the bilayer WC via a kinetic mechanism. Such effects will be important if the defect tunneling barriers are reduced compared to those for direct exchange, as has been found in the monolayer WC \cite{Kim:2022,Kim2024}. In phases II, III, and IV -- where the ground states are primarily VD or IVBS -- the lattices are bipartitie and the VD hopping may be expected to generate Nagaoka-like ferromagnetic interactions \cite{Nagaoka1966}\footnote{In the present situation one complication is that the defect configurations with reduced symmetry will lead to a considerably more complex magnetic Hamiltonian, as the defect now carries an additional ``orbital'' index labeling the inequivalent configurations  \cite{Kim2024} }. These would compete with the antiferromagnetic interactions generated by two and four-particle ring-exchange \cite{esterlis2024}, likely leading to the formation of ferromagnetic polarons. In the IVBS regions of the phase diagram, the dynamics of the intersitital should similarly mediate ferromagnetic interactions in the opposite layer. This is because the smallest closed path involving nearest-neighbor interstitial hopping permutes an odd number of electrons, leading to ferromagnetism \cite{Thouless1965}. The kinetic magnetism in phase V -- where the ID is the ground state -- is likely to be more complex \cite{Kim2024}. 

Interactions between like defects are neglected in this work and require future investigation. These effects may have important consequences, potentially leading to new phases such as superconductivity \cite{Candido2001}.

\begin{acknowledgments}
The authors wish to acknowledge A. Levchenko and D. Zverevich for collobration on related work, and also K. S. Kim and S. A. Kivelson for stimulating discussions. This research was supported by the National Science Foundation (NSF) through the University of Wisconsin Materials Research Science and Engineering Center Grant No. DMR-2309000 (I. E.), NSF Grant No. DMR-2203411 (Z. Z.). Support for this research was also provided by the Office of the Vice Chancellor for Research and Graduate Education at the University of Wisconsin -- Madison with funding from the Wisconsin Alumni Research Foundation, as well as from the University of Wisconsin -- Madison (I. E.).
\end{acknowledgments}
\clearpage
\appendix
\section{Ewald summation technique}\label{app:Ewald}

In this appendix we record the Ewald summation formulas used to compute the defect energies in the bilayer WC. The Ewald summation technique allows one to express the potential energy sums as two parts that quickly converge in real space and momentum space, respectively \cite{Bonsall1977,Goldoni1996}. Utilizing the supercell described in the main text, the total electrostatic energy of $N$ electrons per supercell is given by (we set the electron charge $e=1$):
\begin{equation}
    E_\text{cl}[\{r_{i}\}]=\sum_{i<j}E_\text{pair}(\mathbf{r}_{i}-\mathbf{r}_{j},d_{ij})+\frac{N}{2}E_\text{self}.
\end{equation}
Here $E_\text{pair}$ is the interaction between pairs of different electrons and $E_\text{self}$ denotes the self-interaction between each electron and its image charges. We have
\begin{widetext}
\begin{multline}
        E_\text{pair}(\mathbf{r},d) = \frac{\pi}{\Omega} \sum_{\mathbf{G} \neq 0} \frac{e^{-i \mathbf{G} \cdot \mathbf{r}}}{|\mathbf{G}|}  \left[e^{-|\mathbf{G}|d} \text{erfc}\left(\gamma |\mathbf{G}| - \frac{d}{2\gamma}\right) + e^{|\mathbf{G}|d} \text{erfc}\left(\gamma |\mathbf{G}| + \frac{d}{2\gamma}\right) \right] - \frac{4\sqrt{\pi}\gamma}{\Omega} e^{-d^2/4\gamma^2}\\ + \frac{2\pi d}{\Omega}  \text{erfc}\left(\frac{d}{2\gamma}\right) + \sum_\mathbf{R} \frac{1}{D} \text{erfc}\left(\frac{D}{2\gamma}\right),
\end{multline}

\begin{equation}
    E_\text{self} = \frac{2\pi }{\Omega} \sum_{\mathbf{G} \neq 0} \frac{\text{erfc}(\gamma |\mathbf{G}|)}{|\mathbf{G}|} + \sum_{\mathbf{R} \neq 0} \frac{\text{erfc}(|\mathbf{R}|/2\gamma)}{|\mathbf{R}|} - \frac{4\sqrt{\pi}  \gamma}{\Omega} - \frac{1}{\sqrt{\pi} \gamma}.
\end{equation}
\end{widetext}
In the above equations $D=\sqrt{(\mathbf{r} - \mathbf{R})+d^2}$, $\mathbf{G}$ is the reciprocal vector of the superlattice that satisfies $\mathbf{G}\cdot \mathbf{R}=2\pi n, n\in \mathbb{Z}$, $\Omega=|\mathbf{c}_1\times\mathbf{c}_2|$ is the area of the supercell, and $\gamma$ is the adjustable Ewald parameter. 

The dynamical matrix of an equilibrium configuration is defined by
\begin{equation}
        \hat{C}_{ij}^{\alpha \beta}=\left\{\begin{array}{ll}
      -\partial_\alpha\partial_\beta E_\text{pair}(\mathbf{r},d_{ij})\vert_{\mathbf{r}=\mathbf{r}_{i}-\mathbf{r}_{j}},   & i\neq j \\
        -\sum_{i\neq k} C_{ik}^{\alpha \beta},&  i=j
    \end{array}\right.
\end{equation}
where $\alpha,\beta=x,y$, and
\begin{widetext}
    \begin{multline}
        -\partial_\alpha\partial_\beta E_\text{pair}(\mathbf{r},d)=\frac{\pi}{\Omega}\sum_{\mathbf{G} \neq 0} \frac{G_\alpha G_\beta e^{-i \mathbf{G} \cdot \mathbf{r}}}{|\mathbf{G}|}  \left[e^{-|\mathbf{G}|d} \text{erfc}\left(\gamma |\mathbf{G}| - \frac{d}{2\gamma}\right) + e^{|\mathbf{G}|d} \text{erfc}\left(\gamma |\mathbf{G}| + \frac{d}{2\gamma}\right) \right]\\
        +\sum_\mathbf{R}\left[\text{erfc}\left(\frac{D}{2\gamma}\right)\frac{D^2\delta_{\alpha\beta}-3(\mathbf{r}-\mathbf{R})_\alpha (\mathbf{r}-\mathbf{R})_\beta}{D^5}-\frac{1}{\sqrt{\pi}\gamma}e^{-D^2/4\gamma^2}\frac{3(\mathbf{r}-\mathbf{R})_\alpha (\mathbf{r}-\mathbf{R})_\beta-D^2\delta_{\alpha\beta}+\frac{D^2}{2\gamma^2}(\mathbf{r}-\mathbf{R})_\alpha (\mathbf{r}-\mathbf{R})_\beta}{D^4}\right].
    \end{multline}
\end{widetext}
The eigenvalues $C_k$ of the dynamical matrix $\hat{C}$ give the phonon frequencies $\omega_k$ according to $C_k=m \omega_k^2$. The total zero-point energy is then $E_\text{vib}=\sum_{k=1}^{2N} \hbar\omega_k/2$.

\section{Finite-size scaling of the chemical potential coefficients $B^D$}
\begin{figure}[ht]
    \subfigure[]{\label{fig:scalingiv1}
    \includegraphics[width=0.45 \textwidth]{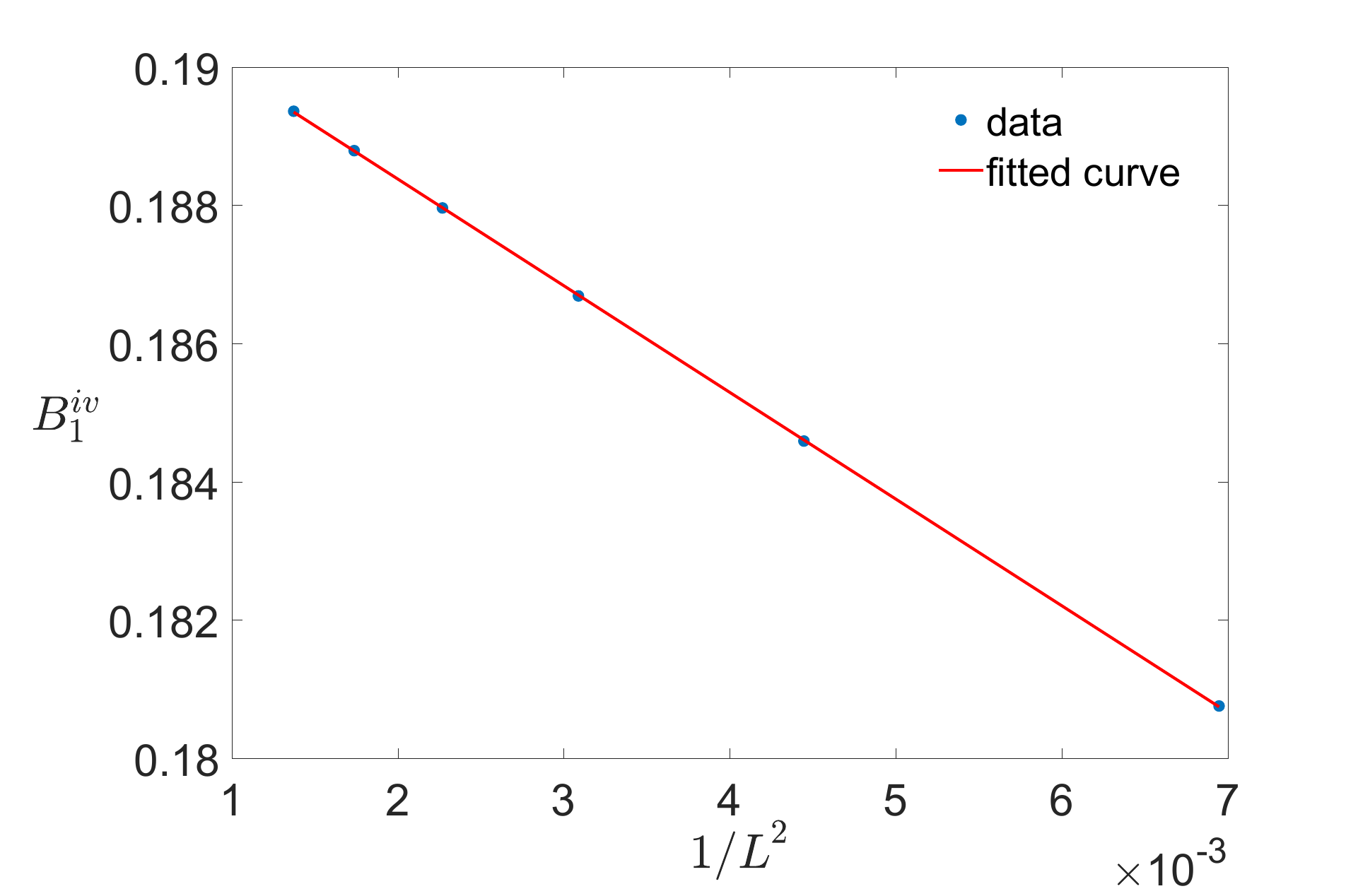}}
    \subfigure[]{\label{fig:scalingiv15}
    \includegraphics[width=0.45 \textwidth]{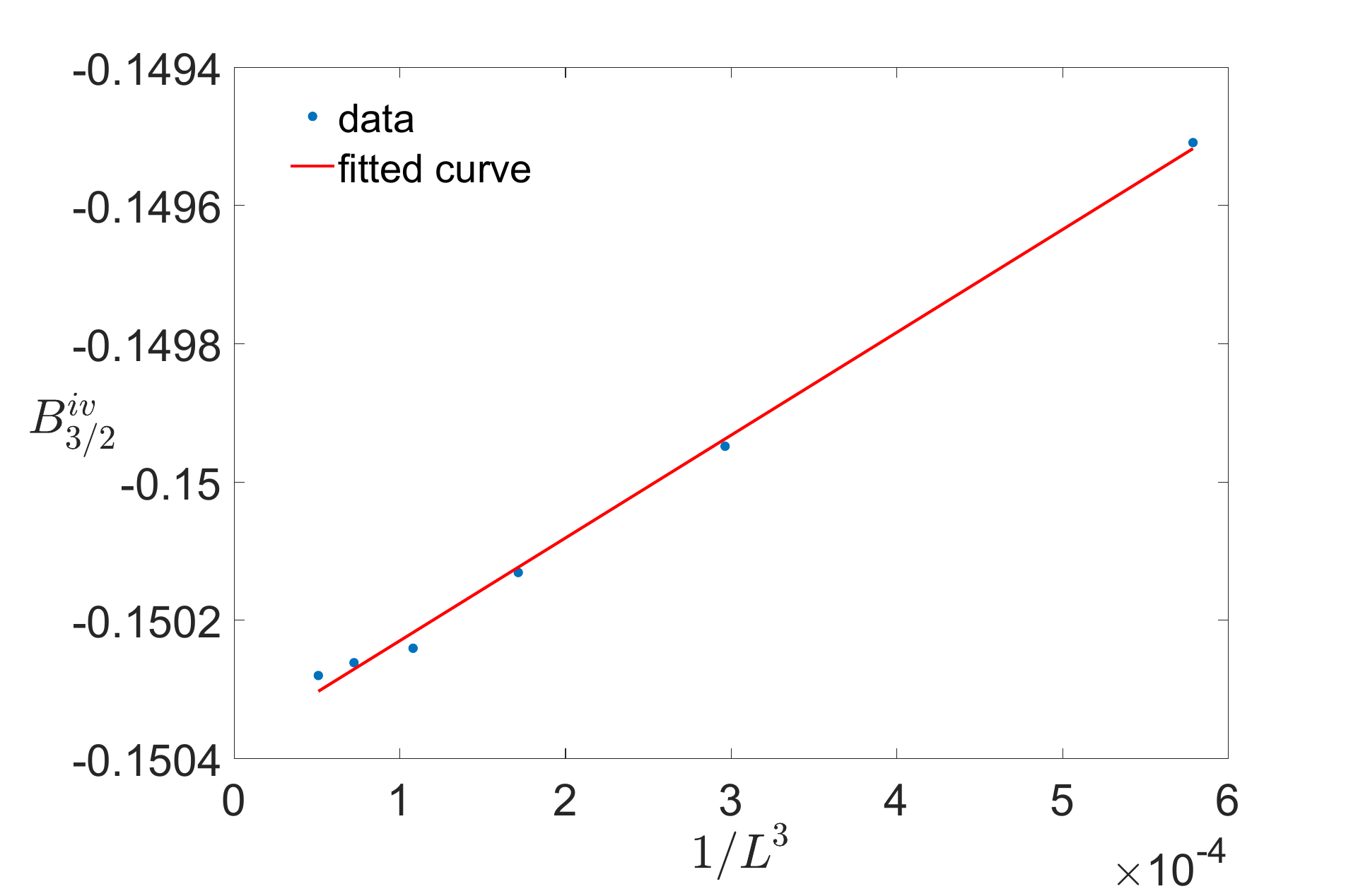}}    
    \caption{The scaling behavior of $B_1^{iv}$ and $B_{3/2}^{iv}$ for $\eta=0.45$.}
   \label{fig:scalingiv} 
\end{figure}

In Figs.~\ref{fig:B1} and \ref{fig:B15} of the main text we report the coefficeints $B_1^{(D)}$ and $B_{3/2}^{(D)}$ extrapolated to the thermodynamic limit $L \to \infty$. (We recall the supercell lattice vectors are taken to be $\mathbf c_{1,2} = L \mathbf a_{1,2}$, where $\mathbf a _1$ and $\mathbf a_2$ are the bilayer WC primitive vectors). As explained in the main text, we have utilized the scaling forms $\Delta B_1^{D} \sim L^{-2}$ and $\Delta B^{(D)}_{3/2} \sim L^{-3}$ for the finite-size corrections, as proposed by Cockayne and Elser \cite{Cockayne1991}. In Fig. \ref{fig:scalingiv} we show representative examples of this scaling.
\bibliography{Ref}
\end{document}